%% file: main_nature_physics.tex
\documentclass[aps,prl,twocolumn,superscriptaddress,nofootinbib,showkeys=false,longbibliography]{revtex4-2}

\usepackage{graphicx}
\usepackage{amsmath,amssymb}
\usepackage{bm}
\usepackage[separate-uncertainty=true,range-phrase=--,range-units=single,per-mode=symbol]{siunitx}
\usepackage{xcolor}
\usepackage[colorlinks=true,linkcolor=blue!60!black,citecolor=blue!60!black,urlcolor=blue!60!black]{hyperref}
\usepackage{booktabs}
\graphicspath{{./}{figs/}{../figs/}}
\input{numbers}

\newif\ifdraft\drafttrue

\newcommand{\ThF}{\ensuremath{\mathrm{ThF}^{+}}}

\DeclareSIUnit{\electron}{e^{-}}
\DeclareSIUnit{\mPa}{mPa}

\begin{document}

\title{Buffer-Gas-Assisted Fluorescence Spectroscopy of Molecular Ions} 

\author{Jinyue Jiang}
\affiliation{Department of Physics, Harvard University, Cambridge, Massachusetts 02138, USA}
\affiliation{Harvard-MIT Center for Ultracold Atoms, Cambridge, MA 02138, USA}
\author{Kenji Shu}
\affiliation{Department of Physics, Harvard University, Cambridge, Massachusetts 02138, USA}
\affiliation{Harvard-MIT Center for Ultracold Atoms, Cambridge, MA 02138, USA}
\affiliation{Fundamental Quantum Science Program (FQSP), RIKEN-TRIP, 2-1 Hirosawa, Wako, Saitama 351-0198, Japan}
\affiliation{Quantum Condensate Research Team, RIKEN Center for Emergent Matter Science (CEMS), 2-1 Hirosawa, Wako, Saitama 351-0198, Japan}
\author{Huaxu Dan}
\affiliation{Department of Physics, Harvard University, Cambridge, Massachusetts 02138, USA}
\affiliation{Harvard-MIT Center for Ultracold Atoms, Cambridge, MA 02138, USA}

\author{Xing Fan}
\email{xingfan@g.harvard.edu}
\affiliation{Department of Physics, Harvard University, Cambridge, Massachusetts 02138, USA}
\affiliation{Harvard-MIT Center for Ultracold Atoms, Cambridge, MA 02138, USA}
\date{\today}

\begin{abstract}
Sustained fluorescence detection of generic molecules has been challenging due to the lack of cycling transitions and limited photon budgets.
We demonstrate a nondestructive method for continuous fluorescence of strongly rovibronically open molecular ions, $^{232}$\ThF. 
Effective quasi-cycling transitions are established by room-temperature buffer-gas collisions, which replenish the population back to the laser-addressed states.
In situ images and rotationally resolved spectra are obtained without optical repumping or auxiliary atomic readout ions.
Combined with low-background spectroscopic detection, this approach achieves a signal-to-noise ratio of $\approx 6$ in a \SI{1}{s} exposure on the $P(1)$ line of $^{232}$\ThF, despite a thermal population of only $\approx 0.17 \%$ in the addressed state at room temperature.
The ability to repeatedly interrogate trapped ions is particularly promising for scarce molecular samples, broadband spectroscopic searches, and in situ monitoring of molecular reactions.
\par
\end{abstract}

\maketitle

\textit{Introduction---}
Direct fluorescence provides a powerful means of continuously probing atomic and molecular structure and providing spatially resolved imaging\cite{demtroder2015laser,zare1974tunable}.
In molecules, spontaneous emission can populate unaddressed rotational, vibrational, or electronic states, limiting the number of photons that can be scattered before the molecules become dark\cite{fitch2021laser,hirota1985spectroscopy,brown2003rotational}.
Repeated optical excitation thus depletes the laser-addressed population, preventing an efficient cycling transition and fluorescence.

Molecules' rich internal structures make them attractive for quantum information\cite{carr2009cold,kaufman2021quantum,demille2002quantum}, precision measurement\cite{cairncross2017precision,Roussy2023,acme2014order,acme2018improved} and chemistry\cite{liu2022bimolecular}.
Only limited favorable molecules support optical cycling even with repump lasers\cite{shuman2010laser,truppe2017molecules,anderegg2017radio,
kozyryev2017sisyphus,vilas2022magneto,hummon2013magneto}. Several spectroscopic methods for generic molecules have been developed using frequency combs, photoionization, photodissociation, quantum-logic techniques, and action spectroscopy\cite{Alighanbari2018,Wolf2016,Chou2017,Wu2026,Gresh2016}.
However, these methods typically generate molecular species at high velocity, require ground-state cooling, or require large samples.
A general spectroscopy method is not widely available and often becomes a bottleneck in surveying new molecular species.

In contrast, buffer-gas collision is an efficient way to access molecules' internal states and redistribute the population\cite{weinstein1998magnetic,maxwell2005high}.
Cryogenic buffer gas is used to create cold molecules and manipulate quantum states\cite{lu2011cold,hutzler2012buffer}.
Recovery of fluorescence by buffer-gas has been demonstrated in atomic and molecular systems ~\cite{HerreraSancho2012,bowen1985quenching,ehrlacher1994quenching,hansen2014rotational,nee2005quenching,notzold2023quenching}. Building on these developments, we exploit collision-induced population recycling to sustain fluorescence from rovibronically open transitions, a method that enables repeated spectroscopic interrogation without optical repumping for generic molecules.

\begin{figure*}[t]
\centering
\includegraphics[width=\textwidth]{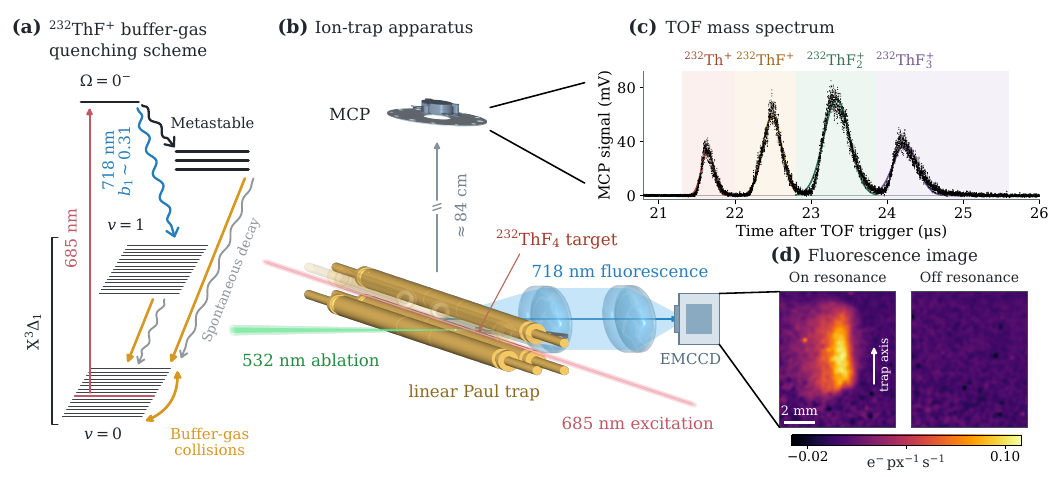}
\caption{%
Buffer-gas-assisted fluorescence detection of trapped \textsuperscript{232}ThF\textsuperscript{+}.
(a) Quasi-cycling scheme for a specific $J''$ level in the $X\,{}^{3}\Delta_{1}(v''=0)\rightarrow[14.59]\,\Omega=0^{-}(v'=0)$ band. Red-shifted fluorescence at 718 nm is collected. Orange arrows indicate buffer-gas-assisted quench from dark rotational, vibrational, and metastable states.
(b) The room-temperature linear Paul trap system. A 532~nm laser ablates a $^{232}$ThF$_4$ target to load $10^{4}$--$10^{5}$ ions per shot. A 685 nm laser addresses a single rotational state. Collection optics image their \SI{718}{nm} fluorescence onto an electron-multiplying CCD (EMCCD). Following optical measurements, ions are extracted toward a micro-channel-plate detector (MCP) for ion number counting.
(c) MCP time-of-flight signal.
(d) Background-subtracted images on resonance and away from resonance.}
\label{fig:setup}
\end{figure*}

\textit{Experiment---}
Figure~\ref{fig:setup}(a) shows the quasi-cycling scheme using $^{232}$\ThF, a heavy ion of particular interest for the electron-electric-dipole-moment (eEDM) study~\cite{Zhou2020,Roussy2023,Ng2022,Gresh2016,ng2025dissociation}.
Through controlled injection, argon (Ar) buffer gas collides with the ions and repopulates the addressed population, enabling direct readout with one continuous-wave excitation laser and no optical repumping.
The ions can be trapped for more than \SI{20}{s} in the current room-temperature Paul trap, allowing long interrogation. 
This nondestructive, \textit{in-situ} readout preserves the ensemble for successive measurements, enabling a frequency scan in a single load.

Figure~\ref{fig:setup}(b) shows the linear segmented Paul trap system with optical access, fluorescence imaging, and time-of-flight (TOF) detection\cite{Schowalter2012,Schneider2014,Schneider2016,Schmid2017}.
A 532 nm pulsed Nd:YAG laser ablates a $^{232}$ThF\textsubscript{4} target \cite{Jonas2025} to produce thorium-containing ions, including \textsuperscript{232}ThF\textsuperscript{+} [Fig~\ref{fig:setup}(c)]. A single ablation shot loads enough \textsuperscript{232}ThF\textsuperscript{+} ions for spectroscopy, typically $10^{4}$--$10^{5}$. The radial and axial secular frequencies are approximately 65 and 10 kHz, respectively. A mass-flow controller supplies Ar to the trap region at pressure up to 330 mPa, providing the collisions that repopulate the laser-addressed state during fluorescence measurements.

A wavemeter-stabilized external-cavity diode laser at 685 nm provides the fluorescence excitation with \SI{7}{mW} maximum power and \SI{1}{mm} radius.
The beam propagates along the trap axis and addresses rotational lines of the $X\,{}^{3}\Delta_{1}(v''=0)\rightarrow[14.59]\,\Omega=0^{-}(v'=0)$ band \cite{Ng2022,Gresh2016}. Collection optics image red-shifted fluorescence near 718~nm onto an EMCCD through a bandpass filter, suppressing scattered excitation light [Fig.~\ref{fig:setup} (d)].
We report background-subtracted photoelectron rates corrected for the camera gain.

Fluorescence is recorded while the ions remain confined, allowing successive measurements at different frequencies before extraction. At the end of the measurement sequence, the ions are extracted by a pulsed kickout field toward a microchannel-plate detector (MCP) for TOF mass spectroscopy shown in [Fig.~\ref{fig:setup}(c)], providing independent ion diagnostics and number normalization.

\begin{figure*}[t]
\centering
\includegraphics[width=\textwidth]{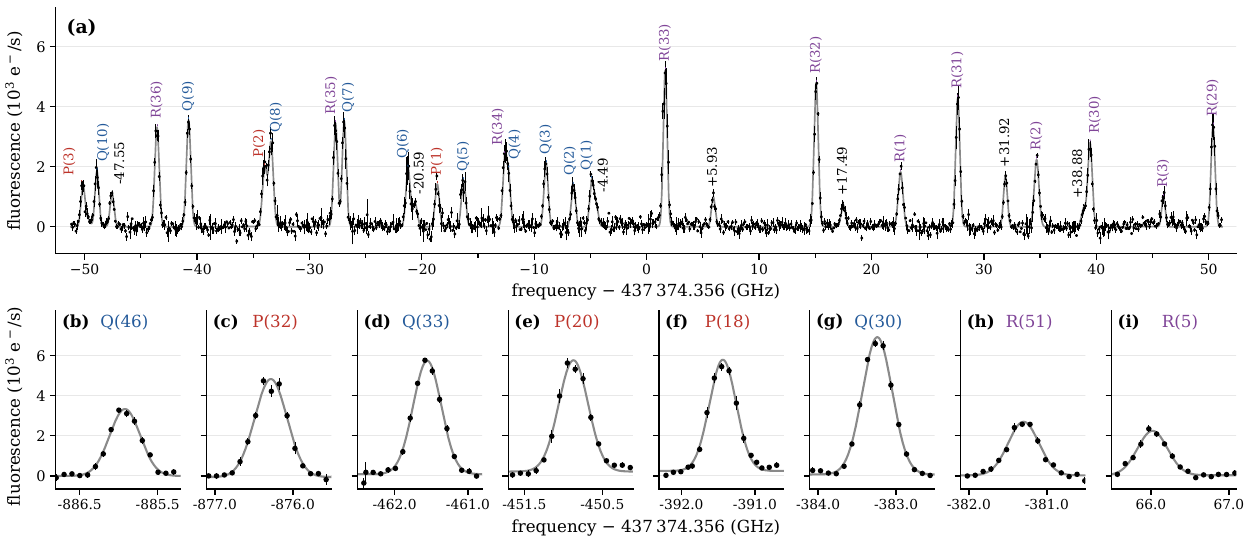}
\caption{%
Fluorescence spectra of trapped ThF\textsuperscript{+} at $\approx$ \SI{120}{mPa} of Ar.
(a) Rotationally resolved spectrum spanning \SI{100}{GHz}.
(b)--(i) Additional frequency windows containing $Q(46)$, $P(32)$, $Q(33)$, $P(20)$, $P(18)$, $Q(30)$, $R(51)$, and $R(5)$.
Black points show photoelectron rates from 1 s exposures, with \SI{100}{MHz} frequency intervals. $P$-, $Q$-, and $R$- branch assignments are labeled in red, blue, and purple. Gray curves show Gaussian profile fits. Black numerical labels in (a) mark unassigned features.
}
\label{fig:spectra}
\end{figure*}

To estimate the photon budget, we use a photon branching fraction $b_{0}\simeq0.62$ into $v''=0$, including the estimated Franck--Condon factors\cite{Ng2024Thesis,Park2026Thesis}.
Without quasi-cycling, an initially addressed ion emits $\langle n_{\gamma}\rangle_{\mathrm{no\,buffer\text{-}gas}}<\frac{1}{1-b_0}\approx2.6$ photons, and a fraction $b_1\simeq0.31$ of them is in the detected 718 nm band ($v'=0\rightarrow v''=1$).
Thus, population replenishment is required to obtain a reusable fluorescence signal on an open rovibronic transition.

Figure~\ref{fig:setup}(d) shows averaged, background-subtracted images of the trapped $^{232}$ThF\textsuperscript{+} cloud. On the $Q(3)$ resonance, the cloud is visible and elongated along the trap axis, with a typical ellipsoidal shape. Away from resonance, no comparable cloud signal is observed even with the same amount of trapped \textsuperscript{232}ThF\textsuperscript{+} ions. 
A typical exposure on resonance yields approximately $1.5\times10^{3}$ signal photoelectrons per second.

To map rotational transitions, we step the excitation frequency by 100 MHz every 1~s during each $\sim$20 s trapping time to cover a rotationally-resolved peak. Each ensemble therefore samples an approximately 1.6 GHz frequency window before extraction. Combining 362 loading cycles yields the survey in [Fig.~\ref{fig:spectra}(a)], spanning $\approx$ \SI{100}{GHz}. In this survey window, we identify 24 transitions in the $P$-, $Q$- and $R$-branches and seven unassigned peaks. The assigned transitions are $P(1)$--$P(3)$, $Q(1)$--$Q(10)$, $R(1)$--$R(3)$, and the returning high-$J$ $R$ branch, $R(29)$--$R(36)$. Additional frequency windows show $Q(46)$, $P(32)$, $Q(33)$, $P(20)$, $P(18)$, $Q(30)$, $R(51)$, and $R(5)$ [Fig.~\ref{fig:spectra}(b)--(i)]. The unassigned peak features are at: \SI{-47.55}{GHz}, \SI{-20.59}{GHz}, \SI{-4.49}{GHz}, \SI{+5.93}{GHz}, \SI{+17.49}{GHz}, \SI{+31.92}{GHz}, and \SI{+38.88}{GHz} from the center. These could be from excitation of other bands of $^{232}$\ThF and are under investigation.

We fit 30 assigned $P$-, $Q$-, and $R$- branch line-centers, extending to $J'' = 51$, using a rotational Hamiltonian that includes centrifugal distortion and lower-state $\Omega$-doubling:
\begin{equation}
\begin{aligned}
\nu_{J',J''} = \nu_0 + c\Bigl[
    & B'J'(J'+1) - D'J'^2(J'+1)^2 \\
    - B''J''(J''+1)  + &D''J''^2(J''+1)^2  + \frac{s_{\mathrm{br}}k''}{2}J''(J''+1)
\Bigr],
\end{aligned}
\label{eq:rotational_fit}
\end{equation}
where $\nu_0$ is the band origin, $c$ is the speed of light, and $J'-J''=-1,0,+1$ for the $P$-, $Q$-, and $R$- branches, respectively.
The branch factor is $s_{\mathrm{br}}=+1$ for $Q$ and
$-1$ for $P$ and $R$, defining our sign convention for
the lower-state $\Omega$-doubling coefficient $k''$.

A simultaneous six-parameter fit to 30 line centers gives an unweighted residual RMS of 14.1~MHz and
$\chi^2/\mathrm{ndf}=24.16/24\approx1.01$. 
The fit most directly determines the differences
$B'-B''=-0.013\,640\,63(50)~\mathrm{cm}^{-1}$ and
$D'-D''=4.39(28)\times10^{-9}~\mathrm{cm}^{-1}$,
where parentheses denote approximate 95\% uncertainties including parameter correlations.
Table~\ref{tab:spectroscopic_constants} compares the fitted constants with Ref.~\cite{Gresh2016}.

\begin{table}[t]
\caption{%
Spectroscopic constants of ${}^{232}\mathrm{ThF}^{+}$ from this work with 95\% confidence level uncertainties in parentheses.
The errors include parameter correlations.
($B''$,\,$D''$, and $k''$: $X^3\Delta_1$ and $B'$ and $D'$: $[14.59]~\Omega=0^-$.)}
\label{tab:spectroscopic_constants}
\centering
\small
\renewcommand{\arraystretch}{1.15}
\setlength{\tabcolsep}{3pt}

\begin{tabular*}{\columnwidth}
{@{\extracolsep{\fill}}lcc@{}}
\toprule
Parameter & This work & Gresh et al.~\cite{Gresh2016} \\
\midrule

$B'\;(\mathrm{cm}^{-1})$
  & $0.228\,998\,5(75)$ & $0.228\,99(2)$ \\

$D'\;(10^{-7}\,\mathrm{cm}^{-1})$
  & $1.352(38)$ & $1.36(2)$ \\

$B''\;(\mathrm{cm}^{-1})$
  & $0.242\,639\,2(78)$ & $0.242\,64(3)$ \\

$D''\;(10^{-7}\,\mathrm{cm}^{-1})$
  & $1.308(41)$ & $1.30(4)$ \\

$|k''|\;(10^{-4}\,\mathrm{cm}^{-1})$
  & $0.8866(38)$ & $0.869(7)$ \\
\bottomrule
\end{tabular*}

\end{table}

\begin{figure*}[t]
\includegraphics[width=\textwidth]{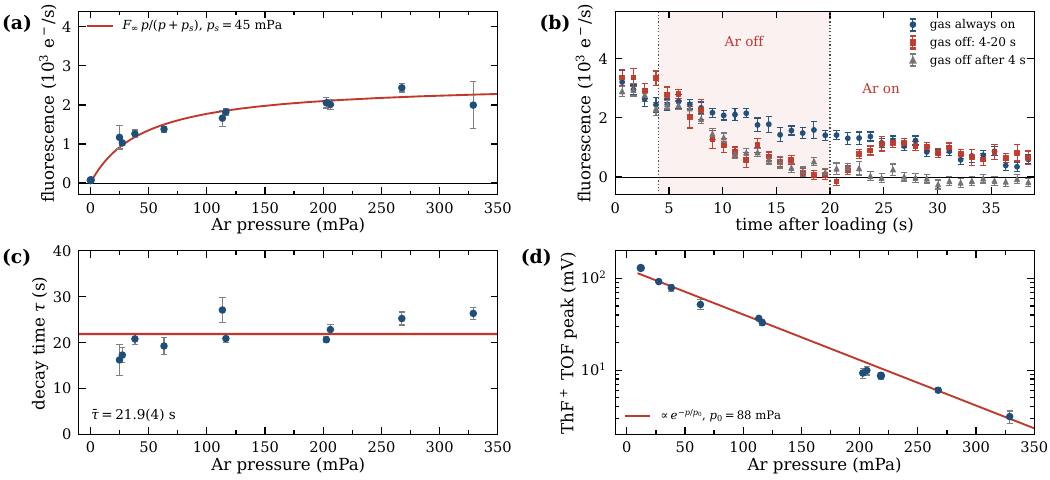}
\caption{%
\label{fig:pressure}
Reversible control of the molecular fluorescence by the buffer gas.
(a)~$Q(3)$ fluorescence versus Ar pressure fitted by $F_{\infty}p/(p+p_{s})$ with $p_{s}=\SI{\pSat(\pSatErr)}{mPa}$.
(b)~Demonstration of the fluorescence induced by buffer-gas: gas on throughout (blue), valve closed \SI{4}{s} after loading (grey), valve closed at \SI{4}{s} and reopened at \SI{20}{s} (red).
(c)~Fluorescence decay time during the hold versus pressure.
(d)~TOF peak height of $^{232}$\ThF\ mass window versus Ar pressure at ejection. Dense buffer gas in the ion extraction path contributes to the signal loss.}
\end{figure*}

Figure~\ref{fig:pressure}(a) shows that the $Q(3)$ signal increases sublinearly over the measured pressure range at full optical power, where an empirical fit $F_{\infty}p/(p+p_s)$ gives $p_s=\SI{\pSat(\pSatErr)}{mPa}$. With the gas flow off, the fluorescence signal vanishes while a substantial TOF signal remains.

To confirm the buffer-gas recycling scheme, we therefore switch the gas flow while holding the same trapped ensemble [Fig.~\ref{fig:pressure}(b)]. Closing the valve 4 s after loading suppresses the fluorescence as the gas is pumped away. Reopening it at 20 s restores the signal without reloading; over 24--38 s, the fluorescence is recovered to the same level, $\approx$ \swRecovery(\swRecoveryErr) times continuous gas cases. Repeats at the 38, 117 and 329 mPa settings show the same behavior.
Fluorescence recovery, together with the open branching of the transition, supports the buffer-gas assisted quasi-cycling interpretation.

Due to chemical reactions induced by collisions with residual background H$_2$O and O$_2$, the fluorescence decreases during each hold, with exponential time constants of $\approx$ \SI{20}{s} [Fig.~\ref{fig:pressure}(c)]. The TOF signals depend on extraction pressure due to collisions with gases during extraction [Fig.~\ref{fig:pressure}(d)]. Therefore, during the spectroscopy, the pressure is regulated to be the same within $0.5\%$ (3$\sigma$) in a single loading event.

\begin{figure*}[t]
\includegraphics[width=\textwidth]{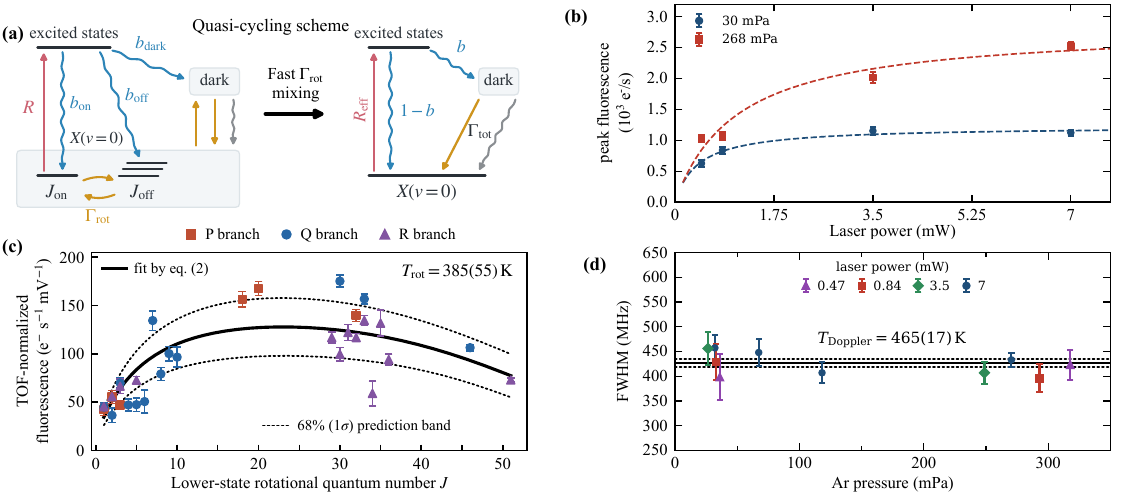}
\caption{%
\label{fig:dyn}
(a)~Quasi-cycling scheme. The laser addresses a $J$-level at excitation rate $R$.
Blue arrows show radiative branching. Orange arrows show population transfer through buffer-gas collisions. The gray arrow shows spontaneous emission from the dark state. Given fast buffer gas mixing between $J$ levels, the scheme could be further simplified to a three-level system.
(b)~Peak fluorescence rate versus laser power at 30 and 268 mPa. Dashed curves are empirical fits of $F_{\max}P/(P+P_{\mathrm{sat}})$. 
(c)~Peak fluorescence rate vs $J$, fitted by the effective three-level system model [eq.~\eqref{eq:return_budget}]. The uncertainties of the fitting parameters are assigned to cover the over-scattering of the data points.
(d)~Gaussian full-width at half-maximum (FWHM) at different pressure and laser powers.}
\end{figure*}

\textit{Fluorescence Model---}
In general, the quasi-cycling scheme can be described as [Fig.~\ref{fig:dyn}(a)]. The bright state is the laser-addressed on resonance $J$ state $J_{\mathrm{on}}$, with dark off resonance $J_{\mathrm{off}}$ in the same $X(v=0)$ manifold, with Doppler broadening also taken into account. The electronic upper state excited by the laser has decay branch $b_{\mathrm{on}}$ back into $J_{\mathrm{on}}$, $b_{\mathrm{off}}$ into unaddressed rotational  states, and $b_{\mathrm{dark}}$ into other dark states, including vibrational dark states, electronic metastable states, etc.

Based on previous works, rotational redistribution rate \(\Gamma_{\mathrm{rot}}\) is of the same order of magnitude as the Langevin collision rate\cite{PerezRios2016Rotational,GonzalezSanchez2018MgH}. An illustrative Langevin collision rate is approximately \(10^4\,\mathrm{s^{-1}}\) at an Ar pressure of \(100\,\mathrm{mPa}\) and a gas temperature of \(300\,\mathrm{K}\). Vibrational quenching and electronic quenching are often on a much slower scale, although their efficiencies depend strongly on the collision partners ~\cite{Stoecklin2016Quenching}. Given fast mixing between rotation states, the model could be simplified into a three-level system as [Fig.~\ref{fig:dyn}(a)].

Let \(R_{\mathrm{eff}}\) denote the effective depopulation rate of the ground manifold due to optical excitation, accounting for suppression by Doppler broadening and the rotational population redistribution rate.
The mean interval between scattered photons is then given by:
\begin{equation}
\frac{1}{\Phi}=\frac{1}{R_{\mathrm{eff}}}+\frac{b}{\Gamma_\mathrm{tot}}.
\label{eq:return_budget}
\end{equation}
Here $\Phi$ is the photon scattering rate per ion, $b$ is the branching ratio into the dark states, and $b_\mathrm{}\Phi$ is the rate of fluorescence that we detect.
Using the quasi-cycling model, the return rate from the dark state is $\Gamma_{\mathrm{tot}}=\Gamma_{\mathrm{spon}}+\Gamma_{\mathrm{gas}}$,
where $\Gamma_{\mathrm{spon}}$ is the spontaneous emission rate. 
Faster population return reduces dark-state delays and supports repeated scattering at higher rates.
Eq.~\eqref{eq:return_budget} gives a joint requirement for the replenishment rate to match the excitation rate: $\Gamma_{\mathrm{tot}} \simeq R_{\mathrm{eff}}b$.
[Fig.~\ref{fig:dyn}(b)] shows that as the pressure increases, both the fluorescence rate and saturation intensity grow, indicating faster repopulation by the buffer-gas.
The saturation against the Ar pressure in [Fig.~\ref{fig:pressure}(a)] suggests that $\Gamma_{\mathrm{tot}}$ with high pressure is dominated by slow quenching, unaffected by the Ar buffer gas.

The three-level system model fits to a rotational temperature of $T_\mathrm{rot}=385\,(55)$~K [Fig.~\ref{fig:dyn}(c)] and is consistent with the temperature from Doppler broadening~[Fig.~\ref{fig:dyn}(d)]. 
The same fit also gives ${\Gamma_{\mathrm{tot}}}/\left(bR_\text{eff}\right)\approx \num{3e-4}$, which indicates that the quenching is still slow compared to the excitation.
By assuming $R_\text{eff} = \qtyrange[retain-unity-mantissa = false, range-phrase = {\text{--}}, per-mode = power]{1E5}{1E6}{\per\s}$ from laser intensity and a typical electric dipole moment of $\sim$1\,Debye, $\Gamma_\mathrm{tot} = $\qtyrange[retain-unity-mantissa = false, range-phrase = {\text{--}}, per-mode = power]{1E1}{1E2}{\per\s}.
This suggests that second-scale metastable states, including vibrational excited state, are efficiently quenched, as well as the rotational states. 

Based on Eq.~\eqref{eq:return_budget}, suitable molecules for quasi-cycling detection should have an accessible optical transition and efficient dark-state returns, either by buffer-gas collisions or by spontaneous emission.
The total off-diagonal fluorescence flux by $N$ trapped ions is
\begin{equation*}
Nb\Phi=Nb
\left[
\frac{1}{R_{\mathrm{eff}}}
+\frac{b}{\Gamma_\mathrm{tot}}
\right]^{-1}
\xrightarrow{R_{\mathrm{eff}}\gg \Gamma_\text{tot}}
N\Gamma_\mathrm{tot}.
\label{eq:quasi_cycling_budget}
\end{equation*}
For example, even for molecules with a branching ratio of $b$ = 0.5 and a metastable state's lifetime of \SI{100}{\milli\s}, assuming a trapped ion number of $N\sim 10^4$ and a collection efficiency $\varepsilon~\sim 1\%$, the demonstrated quench rate in this paper will yield $\varepsilon N\Gamma_\text{tot}=10^{3}$--$10^{4}$ detected photons/s, enough to be observed over a wide range of wavelength.
The estimate here is significantly simplified compared to an actual system with multiple meta-stable states, but it still illustrates the robustness of the method.

\textit{Conclusion---}
By quenching the dark states and mixing the rotational states, buffer-gas-assisted fluorescence spectroscopy enables repeated fluorescence and extended signal accumulation for molecular ions.
Combined with controlled in-trap synthesis, this capability could support spectroscopy of scarce species and monitoring of trapped molecules' reaction.
A further opportunity is comparative spectroscopy of large molecular ions, including aromatic groups with optical cycling centers~\cite{zhu2022functionalizing,augenbraun2022spectroscopy,winnicki2024challenges,mitra2022pathway,zhou2025vibronic} and molecular ions of astrophysical interest~\cite{Campbell2015C60,Campbell2016C70,Hardy2017Coronene,Martin2013Anthracene,Stockett2023CNN}.
The presented method could extend to candidates with substantial vibrational branching without a closed optical cycling or a complete optical repumping scheme.

\textit{Acknowledgments---}
We thank J.~M.~Doyle, D.~DeMille, Y.~Zhou, E.~R.~Hudson, N.~R.~Hutzler, J.~Ye, and E.~A.~Cornell for useful discussions. This work is supported by NSF Grants No.~PHY-2317134 and No.~PHY-2552714, by the RIKEN TRIP (FQSP) and CEMS initiative, and by the Star-Friedman Challenge for Promising Scientific Research at Harvard University.



\textit{Data availability---}
The data supporting the findings of this study are available
from the corresponding author upon reasonable request.


\bibliography{refs}

\end{document}

%% file: numbers.tex
\newcommand{\pSat}{45}
\newcommand{\pSatErr}{7}

\newcommand{\swRecovery}{1.07}
\newcommand{\swRecoveryErr}{9}